\documentstyle[prb,preprint,aps]{revtex}
\begin{document}
\draft

\title{Effects of intrabilayer coupling on the magnetic properties
of YBa$_2$Cu$_3$O$_6$}

\author{Anders W. Sandvik}
\address{National High Magnetic Field Laboratory, Florida State
University, 1800 E. Paul Dirac Dr., Tallahassee, Florida 32306 }

\author{Douglas J. Scalapino}
\address{Department of Physics, University of California,
Santa Barbara, California 93106}

\date{\today}

\maketitle

\begin{abstract}
A two-layer Heisenberg antiferromagnet is studied as a model of the
bilayer cuprate YBa$_2$Cu$_3$O$_6$. Quantum Monte Carlo results
are presented for the temperature dependence of the spin correlation
length, the static structure factor, the magnetic susceptibility, and
the $^{63}$Cu NMR spin-echo decay rate $1/T_{2G}$. As expected, when the
ratio $J_2/J_1$ of the intrabilayer and in-plane coupling strengths is small,
increasing $J_2$ pushes the system deeper inside the renormalized
classical regime. Even for $J_2/J_1$ as small as $0.1$ the correlations are
considerably enhanced at temperatures as high as $T/J_1 \approx 0.4-0.5$.
This has a significant effect on $1/T_{2G}$, and it is suggested that
measurements of this quantity at high temperatures can reveal the strength
of the intrabilayer coupling in YBa$_2$Cu$_3$O$_6$.
\end{abstract}


\vfill\eject

One of the unresolved issues regarding the high-T$_c$ cuprate superconductors
is the role of the coupling between CuO$_2$ planes within the same ``block''
in bi- and tri-layer compounds. Suggestions \cite{altshuler1,mm1}
that this coupling might be responsible for the spin-gap behavior
observed\cite{takigawa} in the bilayer cuprate YBa$_2$Cu$_3$O$_{6.6}$
has spurred recent theoretical work on models of two coupled CuO$_2$
planes.\cite{altshuler2,ubbens,mm2}
Experimentally, the strength of the intrabilayer coupling in
YBa$_2$Cu$_3$O$_{6+x}$ has not yet been accurately determined.
Neutron scattering experiments performed in the insulating regime
have not detected the high-energy branch of the spin-fluctuation
spectrum up to energies of 60meV.\cite{shamoto}
This negative result analyzed in the framework of a linear spin-wave theory
suggests a lower bound of $J_2=8$ meV.\cite{shamoto}
In the recently synthesized high-T$_c$ superconductor
YBa$_4$Cu$_7$O$_{15}$ the two planes constituting a
bilayer have slightly different chemical environments,\cite{stern1}
enabling various cross-relaxation experiments.\cite{rice}
Analyzing such experiments by Stern {\it et al.},\cite{stern2} Millis
and Monien estimated 5 meV $<$ $J_2$ $<$ 20 meV. \cite{mm2} On the other
hand, quantum chemical calculations indicate a value of $J_2$ in
YBa$_2$Cu$_3$O$_{6+x}$ as high as $\approx$50 meV.\cite{barriquand} In this
situation, it is important to consider new possible experiments
that could accurately determine the magnitude of $J_2$.

Here we explore possibilities of extracting the strength of the
intrabilayer coupling from experiments in the antiferromagnetic
insulating regime, e.g. for YBa$_2$Cu$_3$O$_{6}$. We report results of
quantum Monte Carlo (QMC) simulations of a two-layer Heisenberg model
defined by the hamiltonian
\begin{equation}
\hat H = J_1\sum\limits_{a=1,2} \sum\limits_{\langle i,j\rangle}
\vec S_{ai} \cdot \vec S_{aj} + J_2\sum\limits_{i}
\vec S_{1i} \cdot \vec S_{2i},
\label{twolayer}
\end{equation}
where $\vec S_{ai}$ is a spin-$1/2$ operator at site $i$ of
layer $a$, and ${\langle i,j\rangle}$ denotes a pair of
nearest-neighbor sites on a square lattice. This model should be a
reasonable starting point for describing the magnetic properties of
YBa$_2$Cu$_3$O$_{6}$ at temperatures higher than the Neel temperature,
which in clean samples is as high as 500K. Since the in-plane coupling
$J_1 \approx 1200K$,\cite{shamoto} temperatures of interest are
$0.4 \alt T/J_1\alt 0.5$. We present results in this regime for the
spin-spin correlation length, the static structure factor, the
susceptibility, and the $^{63}$Cu NMR spin-echo decay rate $1/T_{2G}$,
and discuss the possibilities to experimentally detect the influence
of a weak $J_2$.

The finite-temperature physics of two-dimensional quantum antiferromagnets was
explained in the framework of a mapping onto a nonlinear $\sigma$-model
some time ago by Chakravarty, Halperin and Nelson.\cite{sigmam}
If the ground state is ordered the correlation length $\xi$ in the
low-temperature renormalized classical (RC) regime
diverges as $e^{2\pi\rho_s/T}$, where $\rho_s$ is the spin-stiffness
constant. At temperatures $T \gg \rho_s$, $T$ is the only relevant
energy scale and the behavior is quantum critical (QC), with
$\xi \sim T^{-1}$.\cite{sigmam,critical} If $\rho_s$ is large, the cross
over boundary between the RC and QC regimes of the $\sigma$-model
may be at a temperature where a $\sigma$-model description of the
antiferromagnet is no longer valid, and instead there is a direct
cross-over from the RC to a local moment (LM) regime where the spins
are essentially decoupled.

For a single Heisenberg plane $(J_2/J_1=0$), the behavior at the
temperatures of interest here is influenced by all the above regimes. For
$0.4\alt T/J_1 \alt 0.6$ there is a cross-over from RC to QC behavior,
and at higher temperatures LM  effects influence the behavior in the
narrow QC regime.\cite{elstner,sandvik1} For the two-layer model described
by (\ref{twolayer}), the ground state order increases with $J_2$ for
small $J_2/J_1$, and has a maximum around $J_2/J_1\approx 0.8$ before
decreasing due to the tendency to singlet formation across the planes for
larger $J_2$.\cite{mm1,chubukov} A quantum phase transition to a disordered
ground state occurs at $(J_2/J_1)_c \approx 2.5$.\cite{hida,sandvik2}
The finite-temperature properties of near-critical systems have been
studied numerically in detail, enabling a direct verification of the
applicability of $1/N$ calculations for the nonlinear $\sigma$-model
in the QC regime.\cite{sandvik2,elstner,sandvik3} Here we are concerned
with values of $J_2/J_1$ more reasonable for modeling high-T$_c$ bilayer
cuprates, and choose $J_2/J_1 = 0.1,0.2$, and $0.5$. Owing to the
enhanced ground state order for this range of $J_2/J_1$, one can expect
the system to exhibit RC behavior at temperatures higher than for a
single layer.\cite{mm1} Below we present quantitative results for several
experimentally accessible quantities.

We begin by defining spin operators that are symmetric and antisymmetric
with respect to interchange of the two layers:
\begin{equation}
\vec S_{\pm}(i)= \vec S_{1i} \pm \vec S_{2i} .
\end{equation}
Using these we have calculated the correlation functions
\begin{equation}
C_{\pm}(\vec r_i - \vec r_i) = \langle S^z_\pm (i)S^z_\pm (j) \rangle,
\end{equation}
the corresponding static structure factors
\begin{equation}
S_\pm (\vec q) = {1\over 2L^2} \sum\limits_{i,j}
\hbox{e}^{i\vec q \cdot (\vec r_i - \vec r_j)} C_\pm (\vec r_i - \vec r_j),
\label{structure}
\end{equation}
and static susceptibilities
\begin{equation}
\chi_\pm (\vec q) = {1\over 2L^2} \sum\limits_{i,j}
\hbox{e}^{i\vec q \cdot (\vec r_i - \vec r_j)}  \int\limits_0^\beta
d\tau  \langle S^z_\pm (i,\tau)S^z_\pm (j,0) \rangle ,
\label{susceptibility}
\end{equation}
where $S^z_\pm (i,\tau)=$e$^{\tau\hat H}S^z_\pm (i)$e$^{-\tau\hat H}$,
and $L$ is the linear size of the system. The normalization in
(\ref{structure}) and (\ref{susceptibility}) has been chosen such
that the standard definitions for a single plane are recovered for
$J_2/J_1 = 0$. For the numerical simulations we have used a QMC method
based on stochastic series expansion \cite{sandvik4} (a generalization
of Handscomb's method \cite{handscomb}), which is free from errors of the
``Trotter approximation'' used in standard methods.\cite{wlmethod}
All results presented here are for 2$\times$64$\times$64 lattices ($L=64$).

In order to extract the correlation length we fit the correlation
function $C_-(r)$ to the nonlinear $\sigma$-model forms discussed
in Ref. \onlinecite{sandvik3}. The results are shown in Fig.~1.
For $J_2/J_1=0$ we find good agreement with the results by
Makivi\'c and Ding.\cite{makivic} The enhancement of the correlations
with increasing $J_2/J_1$ is evident. The correlation length quickly
approaches the size of the lattice as $T$ is lowered below
$T/J_1 \approx 0.4-0.5$, and in order to obtain reliable results for $\xi$
at lower temperatures lattices larger than $L=64$ would have to be used.

The correlation lengths graphed in Fig. 1 exhibit the exponential
growth characteristic of the RC regime, as expected.\cite{mm1} In order to
more quantitatively address the question at what temperature the
RC description is valid for a given $J_2/J_1$ we study the ratio
$S_-(\pi,\pi)/[T\chi_-(\pi,\pi)]$. In a classical system one alway has
$S/(\chi T)=1$ (for any $\vec q$). A characteristic of the quantum
antiferromagnet in the RC regime is that this relation remains
satisfied close to the antiferromagnetic wave-vector.\cite{sigmam}
On the other hand, for a single plane in the QC regime
$S(\pi,\pi)/[T\chi (\pi,\pi)] \approx 1.1$.\cite{critical,sokol,sandvik3}
For a two-layer model in the QC regime, the gapped mode cannot be
neglected if the temperature is of the same order as the gap,
$\Delta \propto \sqrt{J_1J_2}$, and $S(\pi,\pi)/[T\chi (\pi,\pi)]$
will probably differ slightly from the above one-layer $\sigma$-model
prediction. In any case, the RC value for this quantity should be 1,
and is hence useful for determining whether the system is in the RC regime
or not. The results displayed in Fig.~2 clearly indicate that the coupled
planes approach the RC regime considerably faster than a single plane.

The frequency integrated neutron scattering intensity is given by
a combination of the structure factors $S_+(\vec q)$ and $S_-(\vec q)$,
with weights depending on the momentum transfer perpendicular to the
planes.\cite{shamoto} Since the fluctuations corresponding to
$S_+(\vec q)$ are gapped at $\vec q =(\pi,\pi)$, $S_+(\pi,\pi)$ saturates
below a temperature set by the gap, whereas  $S_-(\pi ,\pi)$ diverges.
Fig.~3 shows $J_2/J_1 = 0.1$ results for for $\vec q$ close to
$(\pi,\pi)$. The ratio $S_-(\pi,\pi)/S_+(\pi,\pi)$ is $\approx 4$
already at $T/J_1 = 0.5$, and increases rapidly with decreasing $T$.
At $T/J_1 = 0.4$ the ratio is almost 30 (not shown in the figure). The
smallness of $S_+(\vec q)$ for $\vec q$ close to $(\pi,\pi)$ may make
the detection of the high-energy mode difficult, in particular because
the even and odd modes cannot be completely separated
experimentally.\cite{shamoto}

In Fig.~4 shows results for the uniform susceptibility per spin,
$\chi = \chi_+(q=0)$. The susceptibility for $J_2/J_1 \le 0.2$ is very
close to the single-plane result over the whole temperature range
considered here. The linear behavior seen for the single plane in the
temperature regime $0.35 \alt T/J \alt 0.55$ is in close quantitative
agreement with the prediction for the QC regime,\cite{critical} despite
of the fact that the system actually crosses over to the RC regime at these
temperatures (see Fig.~2).\cite{elstner,sandvik1} An approximately
linear behavior persists at these temperatures also for the systems with
$J_2/J_1 = 0.2$ and $0.5$, which are even deeper inside the RC regime.
It would be interesting to compare the behavior with the QC and RC
predictions for these couplings. As discussed above, this probably
requires calculations for a full two-layer non-linear $\sigma$-model.

We now turn to what we consider the most promising experiment for
determining $J_2$ in the insulating regime. Recently, we
showed \cite{sandvik1}  that the NMR rates $1/T_1$ and $1/T_{2G}$ for
La$_2$CuO$_4$ measured by Imai {\it et al.}\cite{imai} and Matsumura
{\it et al.}\cite{matsumura} are well reproduced within the single-layer
Heisenberg model and known hyperfine form factors (for the
high-temperature regime, similar results were obtained by
Sokol and co-workers \cite{sokol,sokol2}). At high temperatures both
$1/T_1$ and $1/T_{2G}$ show evidence of QC behavior, although the proximity
to the RC and LM regimes influences the behavior as well. As shown above,
even a small intrabilayer coupling pushes the system considerably
further inside the RC regime. As a consequence, a system with
$J_2/J_1 \approx 0.1-0.2$ should exhibit no QC behavior in the experimentally
accessible temperature range. Below we present QMC results for $1/T_{2G}$
which should be useful for direct comparisons with experiments.

The gaussian component of the spin-echo decay rate
is related to the the nuclear spin-spin interactions mediated by the
electronic spins. The coupling of a $^{63}$Cu nuclear spin $\vec I_0$
at site $0$ in plane $a$ to surrounding electronic spins is
approximately given by the Mila-Rice form \cite{mila,mmp}
\begin{equation}
^{63}{\hat H} = A_{\perp} (I^x_0S^x_{a0} + I^y_0S^y_{a0}) +
A_{\parallel} I^z_0S^z_{a0} +B\sum\limits_\delta \vec I_0 \cdot \vec
S_{a\delta} ,
\label{hyperfine}
\end{equation}
where $\delta$ denotes a nearest-neighbor of site $0$. The hyperfine coupling
constants $A_\perp,A_\parallel$ and $B$ are known from Knight shift
measurements.\cite{mmp} Pennington and Slichter derived the following form
for $1/T_{2G}$, expected to be valid for the cuprates \cite{pennington}:
\begin{equation}
{1\over T_{2G}} =
\Bigl [{0.69\over 2\hbar^2} \sum\limits_{\vec x \not= 0}
J_z^2 (0,\vec x) \Bigr ] ^{1/2}.
\label{t2}
\end{equation}
Here $J_z(\vec x_1,\vec x_2)$ is the $z$-component of the induced
interaction between nuclei at $\vec x_1$ and $\vec x_2$,
\begin{equation}
J_z(\vec x_1,\vec x_2) = -\hbox{$1\over 2$}
\sum\limits_{i,j} A(\vec x_1 - \vec r_i)A(\vec x_2 - \vec r_j)
\chi (i-j),
\end{equation}
where for the hyperfine coupling (\ref{hyperfine}) one has
$A(0)=A_\parallel$, $A(1)=B$, and $A(r)=0$ otherwise.
$\chi (i-j)$ is the static susceptibility (\ref{susceptibility}) written
in coordinate space. Note that $\vec x$ in Eq.~(\ref{t2}) runs over
all the spins of both planes,
except the spin at site $0$ in the plane where the nucleus considered
resides. The constant $0.69$ in (\ref{t2}) is the natural
abundance of the $^{63}$Cu isotope.

In Fig.~5 we present results for $J_1/T_{2G}$ versus $T/J_1$ in units
of K/s. The results expected in an experiment can be obtained by dividing
with the relevant value of $J_1$. We have used the standard experimental
values for the hyperfine couplings; $B=41$kOe$/\mu_B$ and
$A_\parallel =-4B$.\cite{mmp} In Ref.~\onlinecite{sandvik2} we have
shown that the QMC results for $J_2=0$ agree well with the
measurements on La$_2$CuO$_4$ \cite{imai,matsumura}
(the best agreement is obtained with a slightly smaller value
for $B$; $B \approx 37$kOe$/\mu_B$). We believe that our
$1/T_{2G}$ QMC results in Fig.~5 will be useful for determining the
value of $J_2$ in YBa$_2$Cu$_3$O$_6$, provided that measurements can be
carried out in the regime of temperatures
500K $\alt T\alt$ 800K. A potential difficulty is that $J_1$ has to
be known to rather high accuracy in order to establish the relation
to the temperature scale of Fig. 5.

In summary, we have studied the effects of a small intrabilayer
coupling on the properties of the two-dimensional Heisenberg model.
Even coupling ratios $J_2/J_1$ as small as 0.1-0.2 push the
boundary of the RC regime up close to the highest temperatures accessible
experimentally. This should have detectable consequences for a number
of quantities. In particular, we suggest that measurements of the
spin-echo decay rate $1/T_{2G}$ at high temperatures would be useful
for determining the strength of the intrabilayer coupling in
YBa$_2$Cu$_3$O$_6$. The spin-lattice relaxation rate, which we have
not yet calculated for the two-layer model, should also be a
sensitive probe.

We would like to thank A.~Chubukov, A.~Millis, and H.~Monien for useful
discussions. Most of the computations were carried out on a
cluster of DEC Alpha 3000/400 workstations at the Supercomputer Computations
Research Institute at Florida State University. This work was
financially supported by the Office of Naval Research under Grant
No. ONR N00014-93-0495, and the Department of Energy
under Grant No. DE-FG03-85ER45197. A.W.S. would like to thank
the Department of Physics at UC Santa Barbara for its hospitality
while part of this work was carried out.

\begin{figure}
FIG.~1. The spin-spin correlation length vs. temperature for various
strengths of the intrabilayer coupling.
\end{figure}

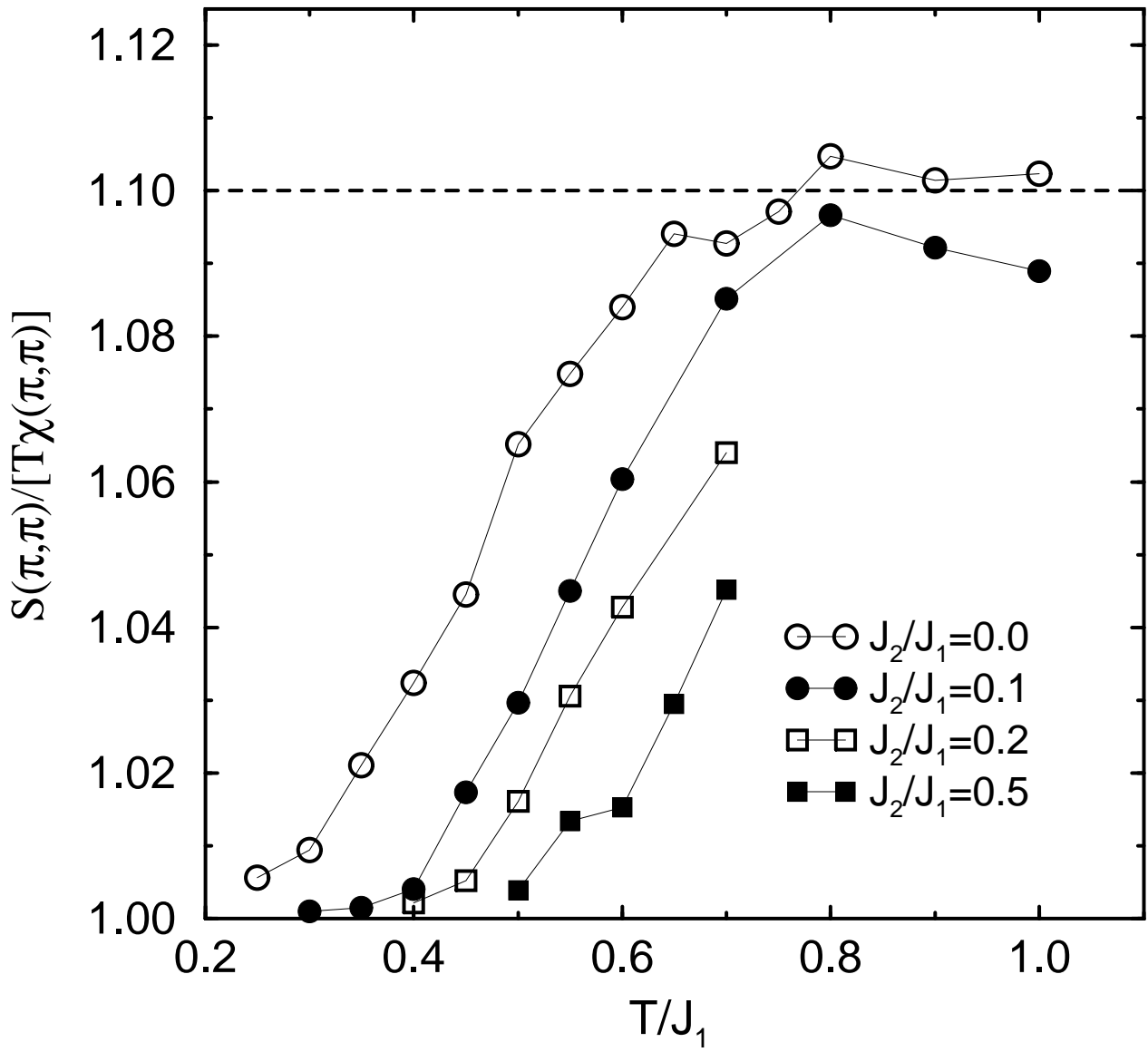
FIG.~2. The ratio $S_-(\pi,\pi)/[T\chi_-(\pi,\pi)]$ vs. $T$ for various
intrabilayer couplings. In the RC regime this ratio is $1$. The
dashed line is the nonlinear $\sigma$-model result for a single plane.

\begin{figure}
FIG.~3. The odd (upper panel) and even (lower panel) static structure factors
for $J_2/J_1 = 0.1$ close to the antiferromagnetic wave-vector
$\vec q=(\pi,\pi)$ at temperatures close to $T/J_1=0.5$.
\end{figure}

\begin{figure}
FIG.~4. The temperature dependence of the uniform susceptibility per
spin for various intrabilayer coupling strengths.
\end{figure}

\begin{figure}
FIg.~5. The $^{63}$Cu spin-echo decay rate $1/T_{2G}$ multiplied
by the in-plane coupling $J_1$ vs. $T/J_1$ for various values of
the intrabilayer coupling strength.
\end{figure}

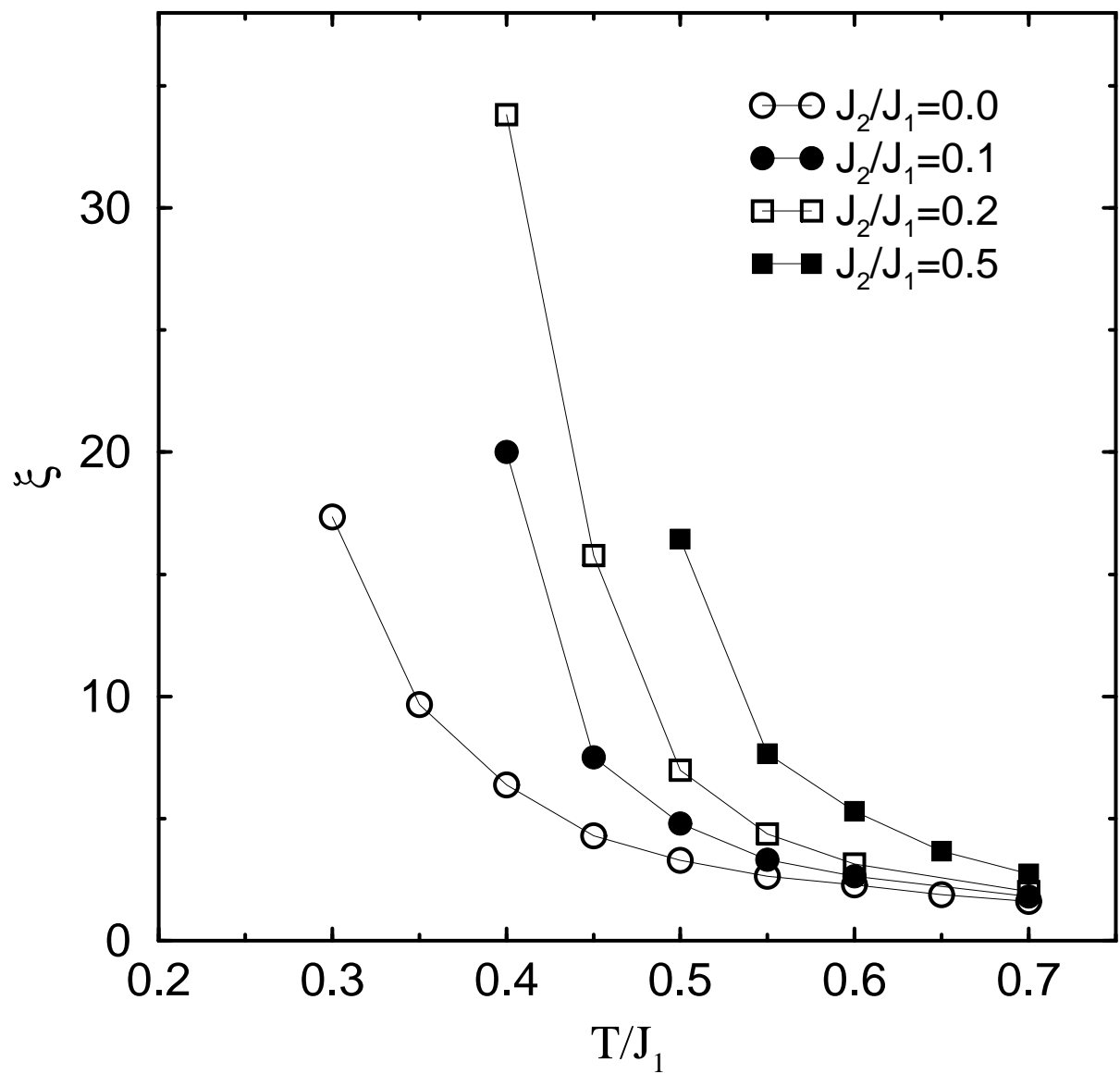
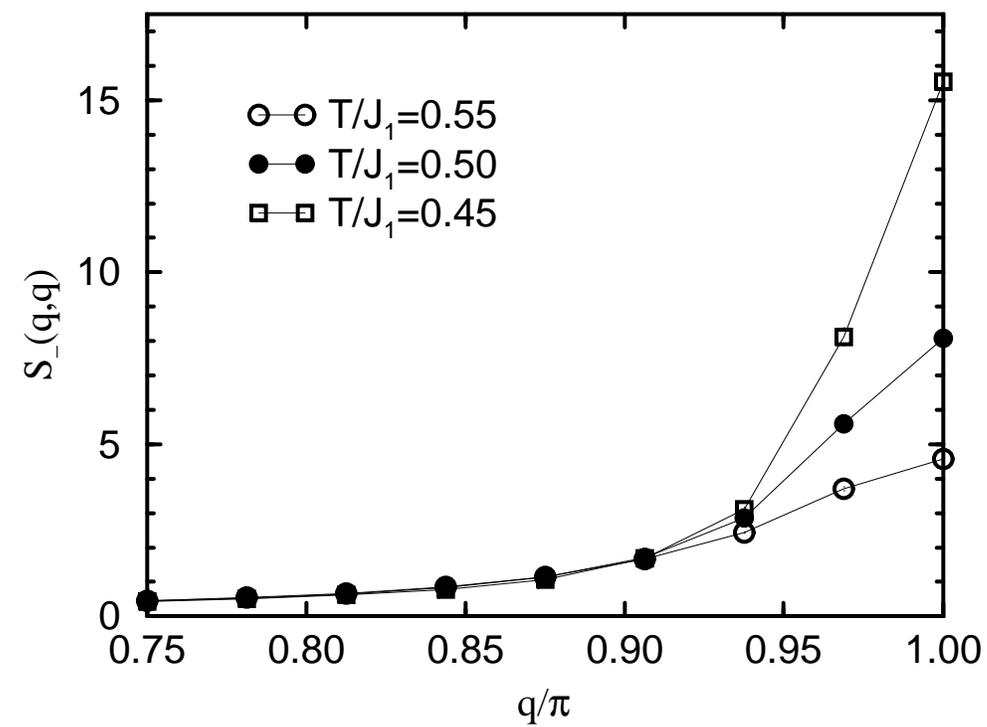
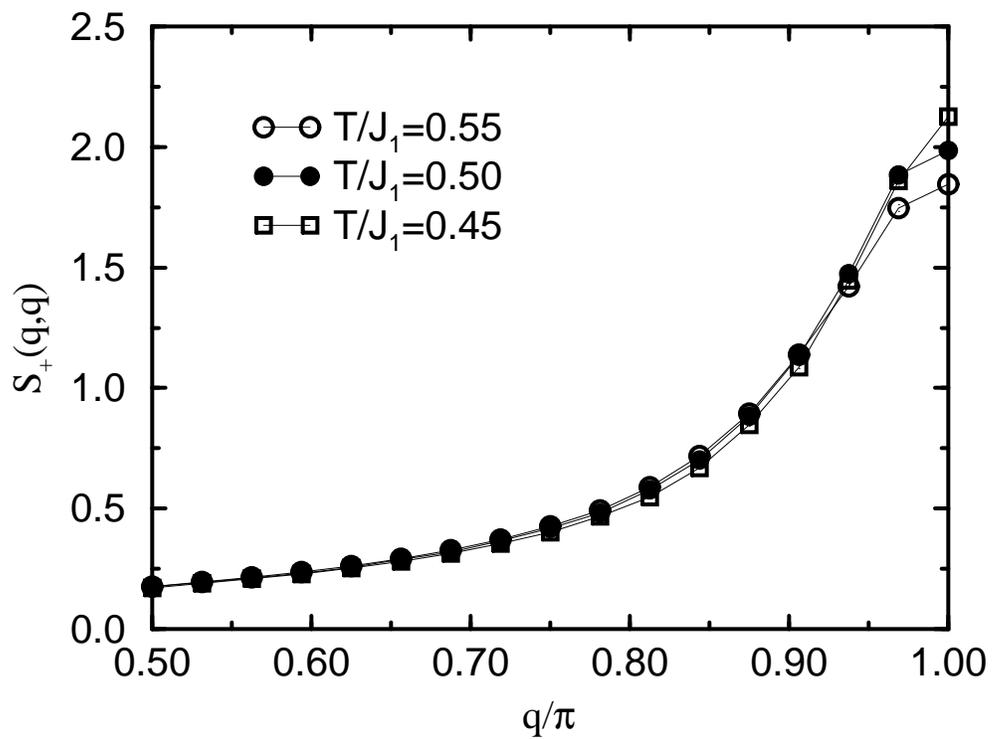
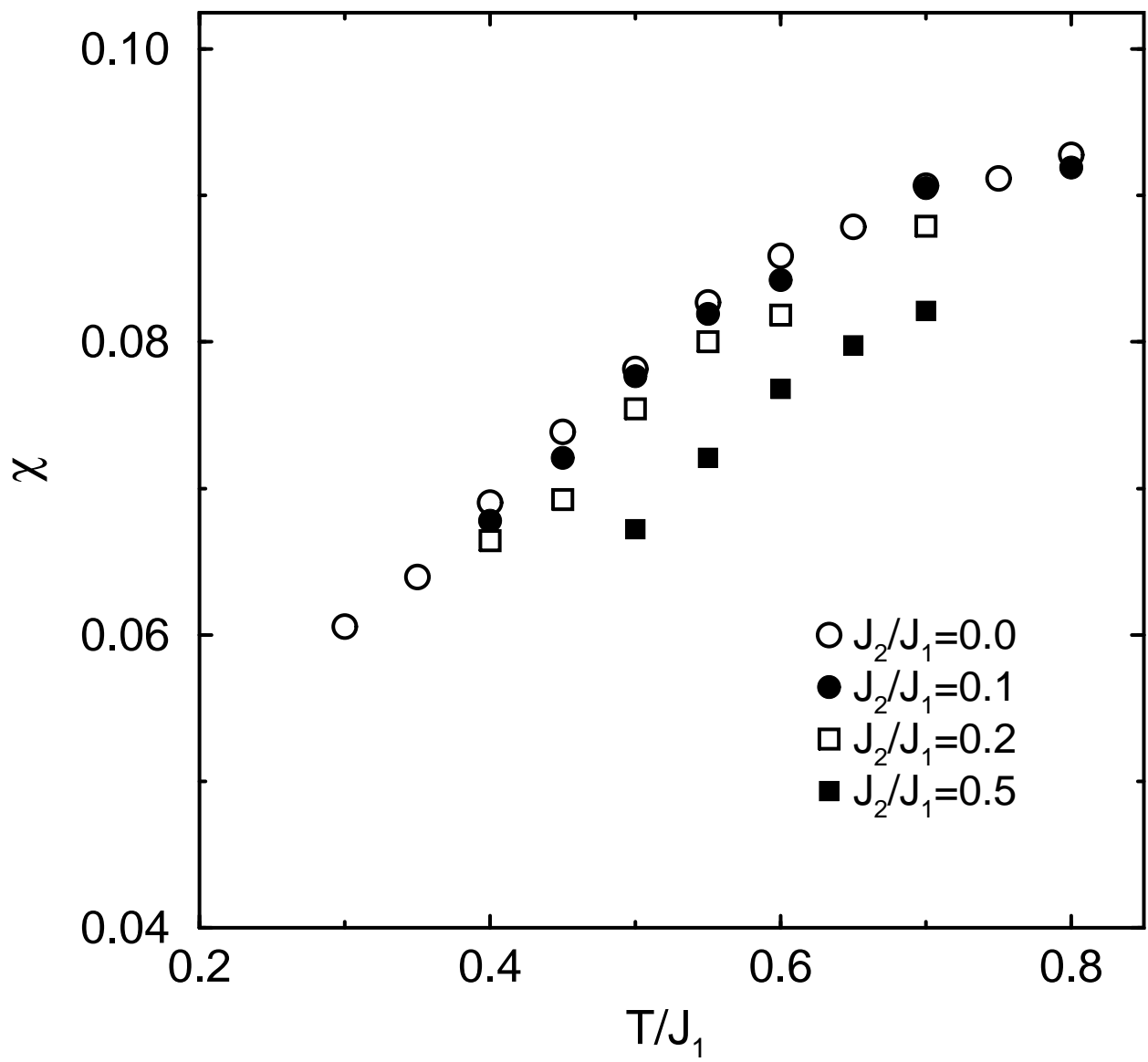
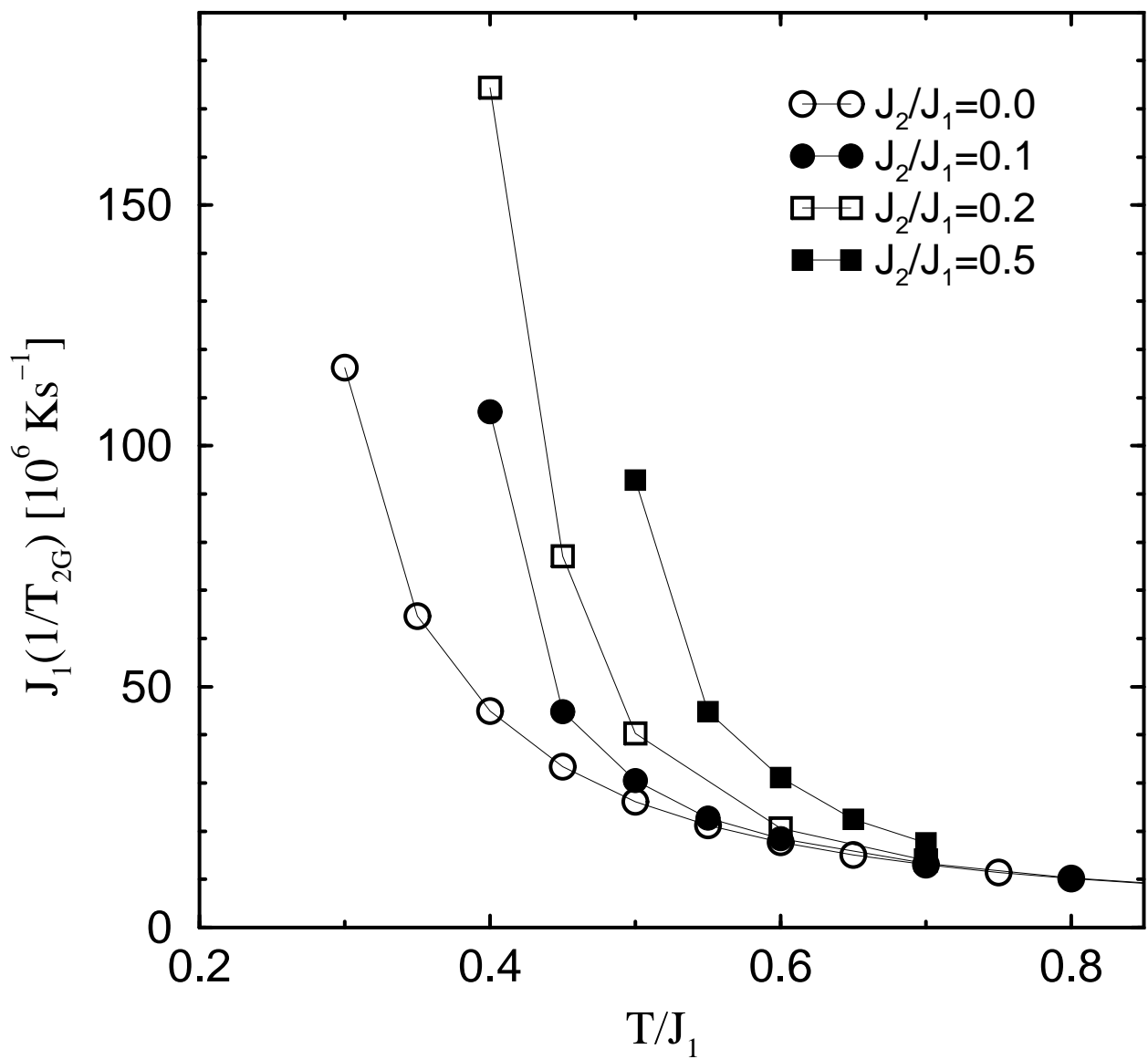
\end{document}